\documentclass[aps,showpacs,nofootinbib]{revtex4}

\usepackage{latexsym}
\usepackage{epsfig}
\usepackage{amssymb}

\newcommand{\lp}{\left(}
\newcommand{\rp}{\right)}

\newcommand{\ba}{\begin{eqnarray}}
\newcommand{\ea}{\end{eqnarray}}
\newcommand{\be}{\begin{equation}}
\newcommand{\ee}{\end{equation}}

\newcommand{\ka}{\kappa}

\newcommand{\R}{\mathcal{R}}

\begin{document}

\title{The Cauchy problem in hybrid metric-Palatini $f(X)$--gravity}

\author{Salvatore Capozziello$^{1.2}$}\email{capozzie@na.infn.it}
\author{Tiberiu Harko$^3$}\email{t.harko@ucl.ac.uk}
\author{Francisco S. N.~Lobo$^{4}$}\email{flobo@cii.fc.ul.pt}
\author{Gonzalo J. Olmo$^{5}$}\email{gonzalo.olmo@csic.es}
\author{Stefano Vignolo$^{6}$}\email{vignolo@diptem.unige.it}

\affiliation{$^1$Dipartimento di Scienze Fisiche, Universit\`{a} di Napoli ``Federico II'', Napoli, Italy
$^2$INFN Sez. di Napoli, Compl. Univ. di Monte S. Angelo, Edificio G, Via Cinthia, I-80126, Napoli, Italy}
\affiliation{$^3$Department of Mathematics, University College London, Gower Street, London
WC1E 6BT, United Kingdom}
\affiliation{$^4$Centro de Astronomia e Astrof\'{\i}sica da Universidade de Lisboa, Campo Grande, Ed. C8 1749-016 Lisboa, Portugal}
\affiliation{$^5$Departamento de F\'{i}sica Te\'{o}rica and IFIC, Centro Mixto Universidad de
Valencia - CSIC. Universidad de Valencia, Burjassot-46100, Valencia, Spain}
\affiliation{$^{6}$DIME Sez. Metodi e Modelli Matematici,
Universit\`a di Genova,  Piazzale Kennedy, Pad. D - 16129 Genova, Italy}

\date{\today}

\begin{abstract}
The well--formulation and the well--posedness of the Cauchy problem is discussed for {\it hybrid metric-Palatini gravity}, a recently proposed modified gravitational theory consisting of adding to the Einstein-Hilbert Lagrangian an $f(R)$ term constructed {\it \`{a} la} Palatini. The theory can be recast as a scalar-tensor one  predicting the existence of  a light long-range scalar field that evades  the local Solar System tests and is able to modify galactic and cosmological dynamics, leading to the late-time cosmic acceleration. In this work, adopting generalized harmonic coordinates, we show that the initial value problem can always be {\it well-formulated} and, furthermore,  can be {\it well--posed} depending on the adopted matter sources. 
 
\end{abstract}


\maketitle
%%%%%%%%%%%%%%%%%%%%%
\section{Introduction}
%%%%%%%%%%%%%%%%%%%%%%%

A central theme in modern cosmology is the puzzling observation  that the Universe seems to be undergoing an accelerated expansion \cite{acc_exp}. The cause of the latter is one of the most important and challenging current problems in cosmology, and remains a tantalizing outstanding question. In fact, the cosmic speed-up represents a new imbalance in the governing gravitational equations, and it is interesting to note that historically, physics has addressed such imbalances by either identifying sources that were previously unaccounted for, or by altering the governing equations. In fact, the standard model of cosmology has favored a missing energy-momentum component in addressing the imbalance, such as the cosmological constant and extensions to dynamical dark energy models (see \cite{Copeland:2006wr} for a review). However, one may consider an alternative approach in considering that General Relativity breaks down at large scales, and a more general action than that of the Einstein-Hilbert action describes the gravitational field. Thus, modified gravity has been an intensive area of research in dealing with the two outstanding problems facing modern cosmology, namely, the late-time cosmic speed-up \cite{CapCurv,Carroll:2003wy} and the dark matter problem \cite{Cap2}. 

Indeed, new features emerge in the latter scenario that may be more successful in providing  covariant infra-red modifications and extensions  of General Relativity. Note that the Einstein field equations of General Relativity were first derived from an action principle by Hilbert, by adopting a linear function of the scalar curvature, $R$, in the gravitational Lagrangian density. Although no {\it a priori} reasons exist to limit the action to this imposition and \cite{early}, in fact, more general gravitational actions involving second order curvature invariants were proposed \cite{Bu70}. The physical motivations for these modifications of gravity were related to the possibility of a more realistic representation of the gravitational fields near curvature singularities and to create some first order approximation for the quantum theory of a gravitational field. In this context, a more general modification of the Einstein-Hilbert gravitational Lagrangian density involving an arbitrary function of the scalar invariant, $f(R)$, was considered \cite{Bu70} in the literature. We refer the reader to Ref. \cite{review} for a recent review on $f(R)$ gravity and to Ref. \cite{Olmo:2011uz} for a review on the Palatini approach to $f(R)$ gravity.
Despite the fact that many of the initially proposed models naturally produced the desired late-time acceleration, it was soon realized that many theories were riddled with serious problems. In particular, taking into account the metric formalism of $f(R)$ gravity, in order to pass the constraints imposed by local tests, it was necessary to impose the existence of a massive scalar field \cite{chameleon}, with an interaction range not exceeding a few millimetres, which obviously, cannot have any impact at large scales. 

In this context, a new class of modified theories of gravity, consisting of the superposition of the metric Einstein-Hilbert Lagrangian with an $f(\R)$ term constructed {\it \`{a} la} Palatini was recently proposed \cite{Harko:2011nh}. It was shown that even if the scalar field is very light, the theory passes the Solar System observational constraints. Therefore the model predicts the existence of a long-range scalar field, modifying the cosmological and galactic dynamics. 
In the latter context, the possibility that the flat rotation curves could be explained within the framework of the recently proposed hybrid metric-Palatini gravitational theory was also explored \cite{Capozziello:2013yha}. The virial theorem was also generalized in the context of the galaxy cluster velocity dispersion profiles predicted by the hybrid metric-Palatini model \cite{Capozziello:2012qt}. Thus, the generalized virial theorem can be an efficient tool in observationally testing the viability of this class of generalized gravity models. The cosmological applications of hybrid metric-Palatini gravity were also explored, and several classes of dynamical cosmological solutions, depending on the functional form of the effective scalar field potential, were explicitly obtained \cite{Hybrid}. 
Furthermore, the cosmological perturbation equations were derived and applied to uncover the nature of the propagating scalar degree of freedom and the signatures of these models in the large-scale structure. In Ref. \cite{Koivisto:2013kwa}, a method was developed to analyse the field content of these theories, in particular to determine whether the propagating degrees of freedom were ghosts or tachyons. In fact, new types of second, fourth and sixth order derivative gravity theories were investigated and the metric-Palatini $f(X)$ theories were singled out as a viable class of ``hybrid'' extensions of General Relativity.

The stability of the Einstein static Universe was also analysed by considering linear homogeneous and inhomogeneous perturbations in the respective dynamically equivalent scalar-tensor representation of hybrid metric-Palatini gravity \cite{Boehmer:2013oxa}. The stability regions of the Einstein static universe were parametrized by the first and second derivatives of the scalar potential, and it was explicitly shown that a large class of stable solutions exists in the respective parameter space. Compact sphere solutions were also recently explored \cite{Capozziello:2012hr}.
The theory was further generalized to include torsion \cite{Capozziello:2013dja}, and a unifying approach was presented where weak forces and neutrino oscillations were interpreted under the same standards of torsional hybrid gravity. This picture allowed the derivation of an effective scalar field which gives rise to a running coupling for Dirac matter fields. In fact, the two phenomena occurring at different energy scales were encompassed under the dynamics of such a single scalar field, which represents the further torsional and curvature degrees of freedom.

However, due to the extra gravitational degrees of freedom emerging from the nonlinearity of the scalar curvature dependence,  which give rise to auxiliary scalar fields, to be a viable theory, the initial value problem needs to be well-formulated and well-posed. In fact, any physical theory is said to be ``physically meaningful'' if an appropriate initial value problem and the boundary conditions are suitably formulated. Thus, starting from suitable initial data on a Cauchy surface, the subsequent dynamical evolution of the physical system is said to be uniquely determined. In this case, it is said that the problem is {\it well-formulated}. Other properties also need to to be satisfied to be a viable theory, such as, if small perturbations occur in the initial data, these need to propagate and produce small perturbations in the subsequent dynamics over the spacetime, where it is defined. The causal structure also needs to be preserved. In fact, a spacetime which possesses a Cauchy surface is said to be {\it globally hyperbolic}, in the sense that the entire future  and past history of the spacetime can be predicted from the initial data on the Cauchy surface \cite{Wald}. Indeed, in a non-globally hyperbolic spacetime, predictability breaks down in that the complete knowledge of the initial conditions do not suffice to determine the entire history of the Universe. If these conditions are satisfied, then the initial value problem of the theory is said to be {\it well-posed}.

It can be shown that General Relativity has a well-formulated and well-posed initial value problem, and as for modified theories of gravity, specific initial value constraints and gauge choices are required in order for the gravitational field equations to be suitable for a correct formulation of the Cauchy problem. Indeed, in the context of $f(R)$-gravity in the metric-affine formalism, by adopting Gaussian normal coordinates, it was shown that the initial value problem is well formulated \cite{Capozziello:2009tea}. In fact, the initial value problem was extensively explored in other approaches of modified gravity, such as $f(R)$ gravity in the presence of perfect fluids \cite{Capozziello:2009pi}, following the well-known Bruhat prescriptions for General Relativity; in the context of metric-affine $f(R)$-gravity in the presence of a Klein-Gordon scalar field acting as a source of the field equations \cite{Capozziello:2010ut}. 
In the metric-affine case, a Hamiltonian description puts forward that the scalar field of the associated scalar-tensor representation is constrained and, therefore, is not a dynamical entity, being an algebraic function of the matter fields. One also finds that the evolution equations of the theory depend, in general, on this non-dynamical scalar and on its {\it spatial} derivatives, but not on its time derivatives \cite{Olmo:2011fh}. This confirms that the initial value problem is well-formulated also for Palatini $f(R)$ theories. The well-posedness of the Cauchy problem was verified for specific forms of matter in \cite{Capozziello:2010ut}.
We refer the reader to \cite{Capozziello:2011gw} for an overview of the initial value problem in $f(R)$ gravity.

Thus, in this work, we consider the Cauchy problem in the hybrid metric-Palatini gravitational theory. This paper is outlined in the following manner: In Section \ref{Sec:review}, we review the hybrid metric--Palatini gravity writing down the field equations and recasting the theory in terms of an equivalent scalar field. In Section \ref{Sec:III}, we analyse the initial value problem in the context of the hybrid metric--Palatini theory, and in Section \ref{Sec:Concl}, we draw our conclusions.

%%%%%%%%%%%%%%%%%%%%%%%
\section{Hybrid metric--Palatini $f(X)$-gravity}\label{Sec:review}
%%%%%%%%%%%%%%%%%%%%%%%

Let us consider the action for hybrid metric--Palatini gravity in $4$-dimensions \cite{Harko:2011nh}, given by
\begin{equation} \label{action}
S= \frac{1}{2\kappa^2}\int d^4 x \sqrt{-g} \left[ R + f(\R)\right] +S_m \ ,
\end{equation}
where in addition to the Einstein-Hilbert term $\sqrt{-g}R$ and the matter action $S_m$, 
which are assumed to have the standard form, there is an extra term depending on both the metric and an independent
dynamical connection $\hat{\Gamma}^\alpha_{\mu\nu}$ through the scalar curvature
\begin{equation}\label{Palatini_curvature}
\R
\equiv  g^{\mu\nu}\R_{\mu\nu} \equiv g^{\mu\nu}\lp
\hat{\Gamma}^\alpha_{\mu\nu , \alpha}
       - \hat{\Gamma}^\alpha_{\mu\alpha , \nu} +
\hat{\Gamma}^\alpha_{\alpha\lambda}\hat{\Gamma}^\lambda_{\mu\nu} -
\hat{\Gamma}^\alpha_{\mu\lambda}\hat{\Gamma}^\lambda_{\alpha\nu}\rp \,,
\end{equation}
where
\begin{equation}\label{Palatini_Ricci}
\R_{\mu\nu} \equiv \hat{\Gamma}^\alpha_{\mu\nu ,\alpha} -
\hat{\Gamma}^\alpha_{\mu\alpha , \nu} +
\hat{\Gamma}^\alpha_{\alpha\lambda}\hat{\Gamma}^\lambda_{\mu\nu}
-\hat{\Gamma}^\alpha_{\mu\lambda}\hat{\Gamma}^\lambda_{\alpha\nu} \,,
\end{equation}
is the Ricci tensor associated with the connection $\hat{\Gamma}^\alpha_{\mu\nu}$. Supposing that the matter Lagrangian is independent of the dynamical connection, variations of the action (\ref{action}) with respect to the metric and the connection yield the field equations
\ba \label{equation1} G_{\mu\nu} +
F(\R)\R_{\mu\nu}-\frac{1}{2}f(\R)g_{\mu\nu} &=& \ka^2 T_{\mu\nu} \,,  \\
%\ee
%and 
%\be 
\hat\nabla_\alpha\/(\sqrt{-g}F(\R)g^{\mu\nu})&=&0\,, \label{equation2}
\ea
respectively, where $F(\R) \equiv df(\R)/d\R$ and $T_{\mu\nu}$ denotes the matter energy-momentum tensor, defined as
 \be \label{memt}
 T_{\mu\nu} \equiv -\frac{2}{\sqrt{-g}} \frac{\delta
 (\sqrt{-g}\mathcal{L}_m)}{\delta(g^{\mu\nu})},
\ee 
and $\ka^2$ is the standard coupling of General Relativity. In view of Eq. (\ref{equation2}), the dynamical connection is compatible with the metric $F(\R)g_{\mu\nu}$, which is conformal to $g_{\mu\nu}$, with the conformal factor  $F(\R)$. This implies the following relation
\ba
\label{ricci} \R_{\mu\nu} & = & R_{\mu\nu} +
\frac{3}{2}\frac{1}{F^2(\R)}F(\R)_{,\mu}F(\R)_{,\nu}
  - \frac{1}{F(\R)}\nabla_\mu F(\R)_{,\nu} -
\frac{1}{2}\frac{1}{F(\R)}g_{\mu\nu}\Box F(\R)\,. \ea 
Replacing Eq. (\ref{ricci}) into the gravitational field equation (\ref{equation1}), the latter can be recast in the form
\ba \label{equation1b}
[1+F(\R)]G_{\mu\nu} & = & \ka^2 T_{\mu\nu} + \nabla_\mu F(\R)_{,\nu} - \Box F(\R)g_{\mu\nu} - \frac{3}{2}\frac{1}{F(\R)}F(\R)_{,\mu}F(\R)_{,\nu} 
    \nonumber \\
&&+ \frac{3}{4}\frac{1}{F(\R)}F(\R)_{,\lambda}F(\R)^{,\lambda}g_{\mu\nu}  - \frac{1}{2}\left(F(\R)\R - f(\R)\right)g_{\mu\nu} \,.
\ea
Moreover, from the trace of the field equation (\ref{equation1}), we have
\be \label{trace} 
F(\R)\R -2f(\R)= \ka^2 T +  R \equiv X\,.
\ee
When Eq. (\ref{trace}) is solvable for $\R$, which provides a suitable expression $\R=\R(X)$ of the Palatini curvature $\R$ as a function of the variable $X$ which measures how much  the theory deviates from the
general relativistic trace equation $R=-\ka^2 T$.

It is possible to show the equivalence between hybrid metric-Palatini $f(X)$-gravity and certain classes of scalar-tensor theories \cite{Harko:2011nh,Hybrid}. To see this point, let us consider the following metric--affine action
\begin{equation} \label{scalar1}
S= \frac{1}{2\kappa^2}\int d^4 x \sqrt{-g} \left[ R + \phi\R-V(\phi)\right] +S_m \,,
\end{equation} 
which represents a metric--affine Brans--Dicke--like theory, with a BD parameter $w=0$ and potential $V(\phi)$ for the scalar field $\phi$. 
Varying this action with respect to the metric, the scalar $\phi$ and the connection lead to the field equations
\begin{eqnarray}
R_{\mu\nu}+\phi \R_{\mu\nu}-\frac{1}{2}\left(R+\phi\R-V\right)g_{\mu\nu}&=&\kappa^2 T_{\mu\nu}\,,
\label{eq:var-gab}\\
\R-\frac{dV}{d\phi}&=&0 \label{eq:var-phi} \,, \\
\hat{\nabla}_\alpha\left(\sqrt{-g}\phi g^{\mu\nu}\right)&=&0  \,, \label{eq:connection}\
\end{eqnarray}
respectively. Again, Eq.~(\ref{eq:connection}) implies that the independent connection is the Levi-Civita connection of
the metric $h_{\mu\nu}=\phi g_{\mu\nu}$. The Ricci tensors $\R_{\mu\nu}$ and $R_{\mu\nu}$ associated respectively with the metrics $h_{\mu\nu}$ and $g_{\mu\nu}$ are then related by
\begin{equation} \label{conformal_Rmn}
%\R=R+\frac{3}{2\phi^2}\partial_\mu \phi \partial^\mu \phi-\frac{3}{\phi}\Box \phi \ ,
\R_{\mu\nu}=R_{\mu\nu}+\frac{3}{2\phi^2}\partial_\mu \phi \partial_\nu \phi-\frac{1}{\phi}\left(\nabla_\mu
\nabla_\nu \phi+\frac{1}{2}g_{\mu\nu}\Box\phi\right) \ .
\end{equation}
Substituting Eq.~(\ref{conformal_Rmn}) into Eq.~(\ref{eq:var-gab}), the latter can be rewritten as
\begin{eqnarray}\label{einstein_phi}
(1+\phi)G_{\mu\nu}&=&\kappa^2T_{\mu\nu} + \nabla_\mu\nabla_\nu\phi - \Box\phi\/g_{\mu\nu} - \frac{3}{2\phi}\nabla_\mu\phi
\nabla_\nu\phi + \frac{3}{4\phi}g_{\mu\nu}\nabla_\lambda\phi\nabla^\lambda\phi - \frac{1}{2}Vg_{\mu\nu}  \label{eq:evol-gab}\, .
\end{eqnarray}
Supposing now that the function $F(\R)=df(\R)/d\R$ is invertible, we can choose the potential $V(\phi)$ of the form
\be\label{potential}
V(\phi)=\phi\/F^{-1}(\phi) - f(F^{-1}(\phi))\,.
\ee
In these circumstances, it is a straighforward matter to verify that Eq. (\ref{eq:var-phi}) amounts to the identity $\phi=F(\R)$ and therefore Eq. (\ref{einstein_phi}) becomes identical to Eq. (\ref{equation1b}).
Moreover, tracing Eq.~(\ref{eq:var-gab}) with $g^{\mu\nu}$,  we find $-
R-\phi\R+2V=\kappa^2T$, and using Eq.~(\ref{eq:var-phi}), we arrive at the following relation
\begin{equation}\label{eq:phi(X)}
2V-\phi \frac{dV}{d\phi}=\kappa^2T+R \ .
\end{equation}
which results to be identical to Eq. (\ref{trace}). Under the hypothesis of Eq. (\ref{potential}), it is then proved that the actions (\ref{action}) and (\ref{scalar1}) are equivalent. Furthermore, if $R$ is replaced in
Eq. (\ref{eq:phi(X)}) with the relation $R=\R+\frac{3}{\phi}\Box
\phi-\frac{3}{2\phi^2}\partial_\mu \phi \partial^\mu \phi$
together with $\R=\frac{dV}{d\phi}$, one then finds that the scalar field $\phi$ is
governed by the second-order evolution equation
\begin{equation}\label{eq:evol-phi}
-\Box\phi+\frac{1}{2\phi}\partial_\mu \phi \partial^\mu
\phi+\frac{\phi[2V-(1+\phi)\frac{dV}{d\phi}]} {3}=\frac{\phi\kappa^2}{3}T\,,
\end{equation}
which is an effective Klein-Gordon equation. 

As a conclusive remark, we notice that Eqs. (\ref{einstein_phi}) and (\ref{eq:evol-phi}) can be derived from the purely metric Brans--Dicke--like action
\begin{equation} \label{scalar2}
S= \frac{1}{2\kappa^2}\int d^4 x \sqrt{-g} \left[ (1+\phi)R +\frac{3}{2\phi}\partial_\mu \phi \partial^\mu \phi
-V(\phi)\right] +S_m \,,
\end{equation}  
with a BD parameter $w=-3/2$. Indeed, the variation of (\ref{scalar2}) with respect the metric tensor gives rise to Eqs. (\ref{einstein_phi}), while the variation with respect to the scalar field yields
\be\label{variation_phi}
R - \frac{3}{\phi}\Box\phi +\frac{3}{2\phi^2}\partial_\mu \phi \partial^\mu \phi - \frac{dV}{d\phi} =0
\ee
Inserting (\ref{variation_phi}) in the trace of (\ref{einstein_phi}), we get exactly (\ref{eq:evol-phi}). The latter shows that the scalar field is dynamical. As we shall see in the next section,
this last fact plays a crucial role in the discussion of the Cauchy problem for hybrid metric-Palatini $f(X)$-gravity, highlighting an important difference with respect to metric--affine $f(R)$-gravity.

%%%%%%%%%%%%%%%%%%%%%%%%%%%%%
\section{The Cauchy problem}\label{Sec:III}
%%%%%%%%%%%%%%%%%%%%%%%%%%%%%%

The dynamical equivalence with scalar--tensor theories shown above is useful to discuss the well--posedness of the Cauchy problem for hybrid $f(X)$-gravity in vacuo and coupled to standard matter sources.
In this perspective, we begin by proving the well--posedness of the Cauchy problem in vacuo, making use of the equivalent formulation (\ref{einstein_phi}) and (\ref{eq:evol-phi}). 
As we shall see, the same conclusions hold in presence of standard matter sources satisfying the usual conservation laws $\nabla^\mu\/T_{\mu\nu}=0\/$. 

Borrowing definitions and notations from \cite{yvonne4}, the key point of our discussion is the introduction of suitable generalized harmonic coorditates, defined by the conditions
\begin{equation}\label{2.2.1}
F^\mu_{\phi}:= F^\mu - H^\mu =0 \qquad {\rm with}\qquad F^\mu :=g^{\alpha\beta}\Gamma^\mu_{\alpha\beta}, \quad H^\mu := \frac{1}{(1+\phi)}\nabla^\mu\/\phi\,.
\end{equation}
As we shall see, the gauge (\ref{2.2.1}) allows us to develop a second order analysis very similar to the one used in GR \cite{yvonne4}. 
We notice that the generalized harmonic gauge (\ref{2.2.1}) is a particular case of the one introduced in \cite{Salgado} to prove the well-posedness of the Cauchy problem for a certain class of scalar-tensor theories of gravity.
 
Let us  start with rewriting  Eqs. (\ref{einstein_phi}) in the form
\begin{equation}\label{2.2.2}
R_{\mu\nu} = \frac{1}{(1+\phi)}\left[\Sigma_{\mu\nu} - \frac{1}{2}\Sigma\/g_{\mu\nu}\right]\,,
\end{equation}
where
\begin{equation}\label{2.2.3}
\Sigma_{\mu\nu}:= \nabla_\mu\nabla_\nu\phi - \Box\phi\/g_{\mu\nu} - \frac{3}{2\phi}\nabla_\mu\phi
\nabla_\nu\phi + \frac{3}{4\phi}\nabla_\lambda\phi\nabla^\lambda\phi - \frac{1}{2}Vg_{\mu\nu}  \,,
\end{equation}
plays the role of an effective energy--momentum tensor. We recall that the  Ricci tensor 
can be expressed as \cite{yvonne4}
\begin{equation}\label{2.2.4}
R_{\mu\nu} = R_{\mu\nu}^\phi + \frac{1}{2}\left[ g_{\mu\sigma}\partial_\nu\left( F^\sigma_\phi + H^\sigma \right) + g_{\nu\sigma}\partial_\mu\left( F^\sigma_\phi + H^\sigma \right)\right]\,,
\end{equation}
with
\begin{equation}\label{2.2.5}
R_{\mu\nu}^\phi := - \frac{1}{2}g^{\alpha\beta}\partial^2_{\alpha\beta}\/g_{\mu\nu} + A_{\mu\nu}\/(g,\partial g)\,,
\end{equation}
where only  first order derivatives  appear in the functions $A_{\mu\nu}$. Due to the assumed gauge condition $F^\mu_\phi =0$ and the explicit expression of $H^\mu$, from (\ref{2.2.4}) and (\ref{2.2.5}),  we get the following representation
\begin{equation}\label{2.2.6}
R_{\mu\nu} = - \frac{1}{2}g^{\alpha\beta}\partial^2_{\alpha\beta}\/g_{\mu\nu} + \frac{1}{(1+\phi)}\partial^2_{\mu\nu}\/\phi + B_{\mu\nu}\/(g,\phi,\partial g,\partial\phi)\,,
\end{equation}
where the functions $B_{\mu\nu}$ depend on the metric $g$, the scalar field $\phi$ and their first order derivatives. At the same time, using Eq. (\ref{eq:evol-phi}) to replace all terms depending on the divergence $g^{\alpha\beta}\nabla_\alpha\nabla_\beta\/\phi$, the right hand side of (\ref{2.2.2}) can be expressed as 
\begin{equation}\label{2.2.7}
\frac{1}{(1+\phi)}\left[\Sigma_{\mu\nu} - \frac{1}{2}\Sigma\/g_{\mu\nu}\right] = \frac{1}{(1+\phi)}\partial^2_{\mu\nu}\/\phi + C_{\mu\nu}\/(g,\phi,\partial g,\partial\phi)\,,
\end{equation}
where, again, the functions $C_{\mu\nu}$ depend only on first order derivatives. A direct comparison of Eq. (\ref{2.2.6}) with Eq. (\ref{2.2.7}) shows  that, in the considered gauge, Eq. (\ref{2.2.2}) assumes the form  
\begin{equation}\label{2.2.8}
g^{\alpha\beta}\partial^2_{\alpha\beta}\/g_{\mu\nu} = D_{\mu\nu}\/(g,\phi,\partial g,\partial\phi)\,.
\end{equation}

The conclusion follows that Eq. (\ref{eq:evol-phi}) together with  Eq. (\ref{2.2.8}), form a quasi-diagonal, 
quasi-linear second-order system of partial differential equations, for which well known theorems by Leray \cite{yvonne4,Leray,Wald} hold. Given initial data on a space-like surface, 
the associated Cauchy problem is then well-posed in suitable Sobolev spaces \cite{yvonne4}. 
Of course, the initial data have to satisfy the gauge conditions $F^i_{\phi}=0$ as well as the Hamiltonian and momentum constraints
\begin{equation}\label{2.2.9}
G^{0\mu}=\frac{1}{(1+\phi)}\/\Sigma^{0\mu} \quad \mu=0,\ldots,3  \,,
\end{equation}
on the initial space-like surface. In connection with this, we notice that, from Eq. (\ref{eq:evol-phi}), 
we can derive the expression of the second partial derivative $\partial^2_0\/\phi$ and replace it on the right hand side of (\ref{2.2.9}), and thus obtaining constraints involving no higher than first order partial derivatives with respect to the time variable $x^0\/$. To conclude, we have to prove that the gauge conditions $F^\mu_\phi =0$ are preserved in a neighborhood of the initial space-like surface. To this end, we first verify that the divergence of the  gravitational field equation (\ref{einstein_phi}) vanishes, namely
\begin{equation}\label{2.2.10}
\nabla^\mu\/\left[(1+\phi)\/G_{\mu\nu} - \Sigma_{\mu\nu}\right] =0 \,.
\end{equation}

Taking into account the identities $\nabla^\mu\/G_{\mu\nu}=0\/$ and $\left(\nabla^\mu\/\phi\right)\/R_{\mu\nu}=\left( \nabla^\mu\nabla_\mu\nabla_\nu - \nabla_\nu\nabla^\mu\nabla_\mu \right)\phi$, automatically satisfied by the Einstein and Ricci tensors, we have
\begin{equation}\label{2.2.11}
\nabla^\mu\/\left[(1+\phi)\/G_{\mu\nu} - \Sigma_{\mu\nu}\right] = -\frac{1}{2}R\nabla_\nu\phi +\nabla^\mu\left( \frac{3}{2\phi}\nabla_\mu\phi\nabla_\nu\phi - \frac{3}{4\phi}\nabla_\lambda\phi\nabla^\lambda\phi\/g_{\mu\nu} + \frac{1}{2}V(\phi)g_{\mu\nu}\right) \,.
\end{equation}
On the other hand, inserting the content of Eq. (\ref{eq:phi(X)}) (in this case, with $T=0$) into the trace of the field equation (\ref{einstein_phi}), we end up with the identity
\be\label{identityR}
R=\frac{dV}{d\phi} + \frac{3}{\phi}\nabla_\lambda\nabla^\lambda\phi - \frac{3}{2\phi^2}\nabla_\lambda\phi\nabla^\lambda\phi \,.
\ee
The identities (\ref{2.2.10}) follow then from a direct comparison of (\ref{2.2.11}) with (\ref{identityR}). 

Now, if $g_{\mu\nu}$ and $\phi$ solve the reduced field Eq. 
(\ref{2.2.8}) and  the scalar field Eq. (\ref{eq:evol-phi}), then we have
\begin{equation}\label{2.2.12}
(1+\phi)\/G^{\mu\nu} - \Sigma^{\mu\nu} = - \frac{(1+\phi)}{2}\/\left( g^{\mu\sigma}\partial_\sigma\/F^\nu_\phi + g^{\nu\sigma}\partial_\sigma\/F^\mu_\phi - g^{\mu\nu}\partial_\sigma\/F^\sigma_\phi \right) \,.
\end{equation}
Identities (\ref{2.2.10}) imply then that the functions $F^\mu_\phi$ satisfy necessarily a linear homogeneous system of wave equations of the form
\begin{equation}\label{2.2.13}
g^{pq}\partial^2_{pq}\/F^i_\varphi + E^{iq}_p\/\partial_q\/F^p_\varphi =0  \,,
\end{equation}
where $E^{iq}_p\/$ are known functions on the space-time. Since the constraints (\ref{2.2.9}) amount to the condition $\partial_0\/F^i_\varphi =0$  on the initial space-like surface, a well known uniqueness theorem for differential systems such as Eq. (\ref{2.2.13}) assures that $F^i_\varphi=0$ in the region where solutions of Eqs. (\ref{eq:evol-phi}) and (\ref{2.2.8}) exist (see also \cite{yvonne4}).

As mentioned above, the illustrated analysis also applies in the case of couplings to standard matter sources such as electromagnetic or Yang-Mills fields, (charged) perfect fluid, (charged) dust, Klein-Gordon scalar fields \cite{Capozziello:2011gw}, so showing the well-posedness of the Cauchy problem for $f(X)$-gravity in presence of standard matter fields. Indeed, when matter sources are present, Eqs. (\ref{eq:evol-phi}) and (\ref{2.2.8}) have to be coupled with the matter field equations. Applying the same arguments developed for GR \cite{yvonne4,yvonne2,yvonne,yvonne3}, it is easily seen that, in the generalized harmonic gauge (\ref{2.2.1}), the matter field equations together with Eqs. (\ref{eq:evol-phi}) and (\ref{2.2.8}) form a Leray hyperbolic and a causal differential system  admitting a well-posed Cauchy problem \cite{Leray}. In addition to the well-known results by Bruhat, the crucial  point is again that the field equations of matter field imply the standard conservation laws $\nabla^\mu\/T_{\mu\nu}=0$. 
This fact allows to verify the validity of Eq. (\ref{2.2.10}) in the presence of matter too ($T_{\mu\nu}\not =0$). We notice that in the considered scalar--tensor theories, the usual conservation laws $\nabla^\mu\/T_{\mu\nu}=0$ have to be necessarily satisfied \cite{Koivisto}.
In summary, the hybrid metric-Palatini gravity satisfies the  well-formulation and well-posedness of Cauchy problem for standard forms of mater and then, in this sense, it is a viable theory.

%%%%%%%%%%%%%%%%%%%%%
\section{Discussion and conclusions}\label{Sec:Concl}
%%%%%%%%%%%%%%%%%%%%%

Hybrid metric-Palatini gravity is a recently proposed modified theory of gravity consisting of adding to the Einstein--Hilbert Lagrangian an $f(R)$ term constructed 
{\it \`{a} la} Palatini. It predicts the existence of a light long--range scalar field that passes the local Solar System tests and is able to modify the galactic and  cosmological dynamics, leading to the late-time cosmic acceleration. Furthermore, several classes of dynamical cosmological solutions were explicitly obtained in this context, and furthermore, the cosmological perturbation equations were derived and applied to uncover the nature of the propagating scalar degree of freedom and the signatures these models predict in the large-scale structure.
However, due to the extra gravitational degrees of freedom emerging from the nonlinearity of the scalar curvature dependence, to be a viable theory, the initial value problem needs to be well-formulated and well-posed. In this context, we considered the Cauchy problem  and showed that the problem is well-formulated and well-posed according to the  Bruhat prescriptions that work for General Relativity. 

Essentially, the demonstration is based on the identification of a generalized  set of harmonic coordinates defined by  suitable (gauge) conditions (\ref{2.2.1}) which allow to rewrite the field equations in the  Einstein-like form (\ref{2.2.2}). Starting from this position, it is possible to reproduce, essentially, the same approach already developed for metric-affine $f(R)$ theories \cite{Capozziello:2009pi}. Finally, the well--formulation and the well-posedness is achieved as soon as viable source matter fields  are considered. In some  sense, this result has to be expected since we are summing up two theories (General Relativity and Palatini $f(R)$ gravity) where the Cauchy problem is self-consistently defined. 

%%%%%%%%%%%%%%%%%%%%%
\section*{Acknowledgments}
%{\it Acknowledgments}.
We thank Tomi Koivisto for enlightening discussions, a careful reading of the manuscript and
for very helpful comments.
SC is supported by INFN (iniziativa specifica QGSKY). FSNL acknowledges financial support of the Funda\c{c}\~{a}o para a Ci\^{e}ncia e Tecnologia through an Investigador FCT Research contract, with reference IF/00859/2012, funded by FCT/MCTES (Portugal), and the FCT grants CERN/FP/123615/2011 and CERN/FP/123618/2011. GJO is supported by the Spanish grant FIS2011-29813-C02-02, the Consolider Programme CPAN (CSD2007-00042), and the JAE-doc program of the Spanish Research Council (CSIC).
%%%%%%%%%%%%%%%%%%%%%

\end{document}